\documentstyle[11pt,astro,newcaption,placeins]{article}

\begin{document}

{\def\RCS#1#2\endRCS{%
  \ifx$#1%
    \XRCS $#2 \endRCS
  \else
    \XRCS $*: #1#2$ \endRCS
  \fi}%
 \def\XRCS $#1: #2,v #3 #4 #5 #6$ \endRCS{%
   \gdef\filename{#2}%
   \gdef\fileversion{v#3}%
   \gdef\filedate{#4}%
   \gdef\docdate{#4}}%
\RCS $Id: cbmlatex.tex,v 1.6 1996/07/14 00:52:17 cmoore Exp $ \endRCS}%
\typeout{*** CBM latex \docdate\space\fileversion\space ***}


\hyphenation{man-u-script man-u-scripts ap-pen-dix data-base data-bases
   an-a-lyse an-a-lyses an-a-lysed an-a-lysing dis-trib-ute dis-trib-uted
   dis-trib-utor dis-trib-utors be-hav-iour be-hav-iours pre-am-ble
   pre-am-bles post-am-ble post-am-bles flocci-nauci-nihili-pili-fi-ca-tion
   fest-schrift Sprach-wissen-schaft skinny-dip skinny-dip-ping skinny-dip-per
   skinny-dip-pers phoe-nix sphyg-mo-mano-meter medi-ocre oeuvres
   demi-semi-quaver hemi-demi-semi-quaver semi-quaver praise-worthy re-qui-site
   pre-requi-site pre-requi-sites re-qui-si-tion pa-ram-e-ter-ised
   poly-go-ni-sa-tion kwashi-or-kor poly-sty-rene poly-sor-bate
   poly-un-sat-u-rate poly-un-sat-u-rated un-nil-quad-ium un-nil-pent-ium
   un-nil-sept-ium un-nil-oct-ium un-nil-enn-ium ein-stein-ium men-del-ev-ium
   ga-do-lin-ium Brit-ish Brit-ain pseudo-intel-lec-tual pseudo-intel-lec-tuals
   re-edu-cate re-edu-cated re-edu-ca-tion re-edu-cat-ing re-edu-cates
   re-en-list re-en-listed re-en-lists re-en-list-ing re-en-list-ment
   Min-ne-apo-lis Mont-real arc-tan-gent arc-co-sine arc-co-tan-gent
   arc-co-se-cant ac-cu-sa-tive Der-zhan-ski Chiang Sing-a-pore Liecht-en-stein
   Czecho-slo-va-kia Kor-do-fan-ian Yu-go-sla-via Ethi-o-pia Bang-la-desh
   Bang-kok Ryd-berg Xiao-ping Flor-ence Flor-en-tine Eri-trea Eri-trean
   Eri-treans Alex-an-dria Ala-bama Ala-ba-man Ala-ba-mans Arkan-sas Arkan-san
   Arkan-sans Colo-rado Colo-radan Colo-radans Del-a-ware Del-a-war-ean
   Del-a-war-eans Car-o-lina Car-o-lin-ian Car-o-lin-ians Da-kota Da-ko-tan
   Da-ko-tans Flor-ida Flor-id-ian Flor-id-ians Kan-sas Kan-san Kan-sans
   Loui-si-ana Loui-si-anan Loui-si-anans Mas-sa-chu-setts Mas-sa-chu-sett
   Min-ne-sota Min-ne-so-tan Min-ne-so-tans Ne-bras-kan Ne-bras-kans Ne-vada
   Ne-vadan Ne-vadans Okla-homa Okla-ho-man Okla-ho-mans Ten-nes-see
   Ten-nes-sean Ten-nes-seans Wis-con-sin-ite Wis-con-sin-ites Wyo-ming
   Wyo-ming-ite Wyo-ming-ites Co-lum-bia Co-lum-bian Co-lum-bians
   Sas-katch-e-wan Sas-ka-toon Win-ni-peg To-ronto Yellow-knife Bruns-wick
   Lab-ra-dor Roch-es-ter Chi-cago Mont-gom-ery Nash-ville Louis-ville
   Al-bu-quer-que In-di-ana-polis dys-lexia anae-mia anae-mic ec-ze-ma
   ec-ze-ma-tous oe-de-ma oe-de-ma-ta rar-e-fac-tion ono-mat-o-poe-ia
   ono-mat-o-poe-ic ono-mat-o-po-et-ic ono-mat-o-po-et-i-cally
   cryp-to-crys-tal-line cryp-to-gram cryp-to-grams zzx-jo-anw Alex-an-der
   politico-economic saw-horse saw-horses Nie-tzsche Nie-tzschean tow-hee
   tow-hees tri-bromo-ethanol para-digm para-digms ser-geant ser-geants
   tribes-man tribes-men tribes-woman tribes-women char-treuse veri-simil-i-tude
   veri-simil-i-tu-di-nous radio-iso-tope radio-iso-topes asa-foet-ida
   asa-fet-ida ra-gout prov-ince prov-inces pro-vin-cial ama-ret-to
   anom-a-lous counter-rev-o-lu-tion counter-rev-o-lu-tion-ary
   counter-rev-o-lu-tion-aries Brook-haven super-con-duc-tor
   super-con-duc-tive super-con-duc-ting super-con-duc-tiv-ity
   super-col-lider sup-col-liders cei-lidh cei-lidh-ean apo-phthegm
   apo-phthegms apo-phtheg-matic apo-phtheg-mat-ical apo-phtheg-mat-i-cally
   electro-encephalo-gram electro-encephalo-grams electro-encephalo-graph
   electro-encephalo-graphs electro-encephalo-graphic electro-encephalo-graphy
   wheel-chair wheel-chairs Pat-rick acu-punc-ture acu-punc-tur-ist
   hippo-pot-a-mus mono-sodium
     anom-a-ly anom-a-lies an-tin-o-my an-tin-o-mies band-leader
   band-leaders bor-no-log-i-cal Brown-ian buzz-word buzz-words cart-wheel
   cart-wheels cho-les-teric data-path data-paths demos Dijk-stra
   electro-mechan-ical electro-mechano-acoustic equi-vari-ance equi-vari-ant
   Euler-ian fermi-ons flow-chart flow-charts Gauss-ian geo-met-ric Greifs-wald
   Grothen-dieck Grund-leh-ren Hamil-ton-ian Her-mit-ian in-fra-struc-ture
   je-re-mi-ads Kadom-tsev Le-gendre Lip-schitz Lip-schitz-ian
   macro-eco-nomic macro-eco-nomics Markov-ian meta-lan-guage meta-lan-guages
   micro-eco-nomic micro-eco-nomics micro-fiche mod-el-ling mono-en-er-getic
   mono-pole mono-strofic mul-ti-plic-able mul-ti-plic-ably neo-fields
   Noether-ian non-emer-gency non-equi-vari-ance non-euclid-ean non-smooth
   pa-ram-e-ter-ized para-mil-i-tary poly-ene poly-go-ni-za-tion
   pseudo-dif-fer-en-tial pseudo-fi-nite pseudo-fi-nite-ly pseudo-forces
   pseudo-word pseudo-words qua-drat-ic qua-drat-ics quasi-equiv-a-lence
   quasi-hy-po-nor-mal quasi-smooth quasi-sta-tion-ary Rie-mann-ian
   sched-ul-ing Schwarz-schild semi-def-in-ite semi-ho-mo-thet-ies
   ser-vo-mech-anism set-up so-le-noid so-le-noids spher-oid spher-oids
   sto-chas-tic sub-scrib-er sub-scrib-ers sum-ma-ble thermo-elas-tic
   time-stamp time-stamps ver-all-ge-mei-ner-te 
   wave-guide}

\def\etal{{\it et al.\/}}
\newcommand{\prhchi}{PRH--$\chi^2$}

\def\deg{\ifmmode^\circ\else$^\circ$\fi}
\def\degC{\ifmmode\,^{\circ}{\rm C}\else$\,^{\circ}{\rm C}$\fi}
\def\Mpc{\ifmmode\,{\rm Mpc}\else$\,{\rm Mpc}$\fi}
\def\GHz{\ifmmode\,{\rm GHz}\else$\,{\rm GHz}$\fi}
\def\MHz{\ifmmode\,{\rm MHz}\else$\,{\rm MHz}$\fi}
\def\Hz{\ifmmode\,{\rm Hz}\else$\,{\rm Hz}$\fi}
\newcommand{\unit}[1]{\ifmmode\,{\rm #1}\else$\,{\rm #1}$\fi}

\def\Msun{\ifmmode\,{\rm M_\odot}\else$\,{\rm M_\odot}$\fi}
\def\Rsun{\ifmmode\,{\rm R_\odot}\else$\,{\rm R_\odot}$\fi}
\def\Mearth{\ifmmode\,{\rm M_\oplus}\else$\,{\rm M_\oplus}$\fi}
\def\arcsec{\ifmmode^{\prime\prime}\else$^{\prime\prime}$\fi}
\def\arcmin{\ifmmode^{\prime}\else$^{\prime}$\fi}
\def\Hi{{\rm\sc Hi}}
\def\Hii{{\rm\sc Hii}}

\def\gtsim{\mathrel{\mathpalette\vereqtwo>}}
\def\ltsim{\mathrel{\mathpalette\vereqtwo<}}
\def\vereqtwo#1#2{\lower3.5pt\vbox{\baselineskip 5.5pt plus 0pt minus0pt\lineskip-.5pt\ialign{$#1\hfil##\hfil$\crcr#2\crcr\sim\crcr}}}

\def\VLAfootnote{\footnote{The VLA is part of the National Radio
Astronomy Observatory, which is operated by Associated
Universities,~Inc., under cooperative agreement with the
National Science Foundation.}} 

\def\MDMfootnote{\footnote{Observations reported here were carried out
at the MDM Observatory on Kitt Peak, which is operated by a three
university consortium consisting of the Univeristy of Michigan,
Dartmouth College, and M.I.T.}}

\def\GBfootnote{\footnote{The Green Bank 140-foot telescope is part of
the National Radio Astronomy Observatory, which is operated by
Associated Universities,~Inc., under cooperative agreement with the
National Science Foundation.}}

\input epsf
\thispagestyle{empty}
\begin{center}
\vspace*{0.1in}
{\large\bf 15\unit{GHz} Monitoring of the Gravitational Lens
MG~0414+0534}\\
\vspace*{0.25in}
{\sc Christopher B. Moore\footnote{Current Address: Kapteyn Astronomical Institute, Postbus 800, 9700~AV~Groningen, The~Netherlands; cmoore@astro.rug.nl} and Jacqueline N. Hewitt\footnote{MIT
    Room 26-327, 77 Massachusetts Avenue, Cambridge, MA 02139; jhewitt@maggie.mit.edu}\\}
 M.I.T. Department of Physics and Research Laboratory for Electronics\\
\bigskip
{\em to appear in the Astrophysical Journal, November 1997}
\vspace*{\fill}
\begin{abstract}

We report the results of monitoring the four images of the
gravitational lens MG 0414+0534 at 15\unit{GHz}.  In 35 VLA maps
spanning 180\unit{days}, we measure root-mean-square variations in the
image light curves of $\sim\!\!3.5\%$ mostly due to variations in the
flux density calibration.  The flux ratios, which are independent of
flux density calibration variations, show root-mean-square variability
of 1--3\%.  Extensive simulations of the data analysis process show
that the observed variations in the flux ratios are likely to be due
entirely to errors in the deconvolution process.  It is possible that
some of the observed variation is due to the source; however, the
signal-to-noise ratio is too small to make a time delay determination
using a data set of this size.

\end{abstract}
\end{center}
\vspace*{.5in}
\vspace*{\fill}
{\tt 
Subject headings: cosmology: observations - 
gravitational lensing -\\ quasars: individual (MG~0414+0534)}

\section{Introduction}

Well before the first gravitational lens was discovered, Refsdal
\cite*{refsdal64a,refsdal64b} pointed out that, given a model of the
lensing potential and the redshift of both the source and the lens, a
measurement of the difference in optical path length between two
images in a gravitational lens could be used to determine the Hubble
parameter.  More recently, Narayan \cite*{narayan91} has shown that a
time delay measurement combined with a model of the lensing potential
provides a measure of the angular diameter distance to the lens that
is independent of the redshift of the source and cosmological
assumptions other than local isotropy and homogeneity transverse to
the line of sight.  Thus, lens models combined with time delay
measurements and lens redshifts can be used to measure the
relationship between angular diameter distance and redshift.  The
measurement of the angular diameter distance--redshift relation at a
variety of lens redshifts will allow the comparison of the measured
relation to that predicted by cosmological models.  Large-scale
structure may contribute to the deflection and time delays in strong
lensing \cite{seljak94a,bar-kana96a} raising the interesting
possibility of detecting the effect of large-scale structure in a
sample of lenses.

\begin{figure}[htb]
\begin{center}
  \epsfxsize=\textwidth
  \begin{center}\leavevmode
   \epsfbox[54 162 592 639]{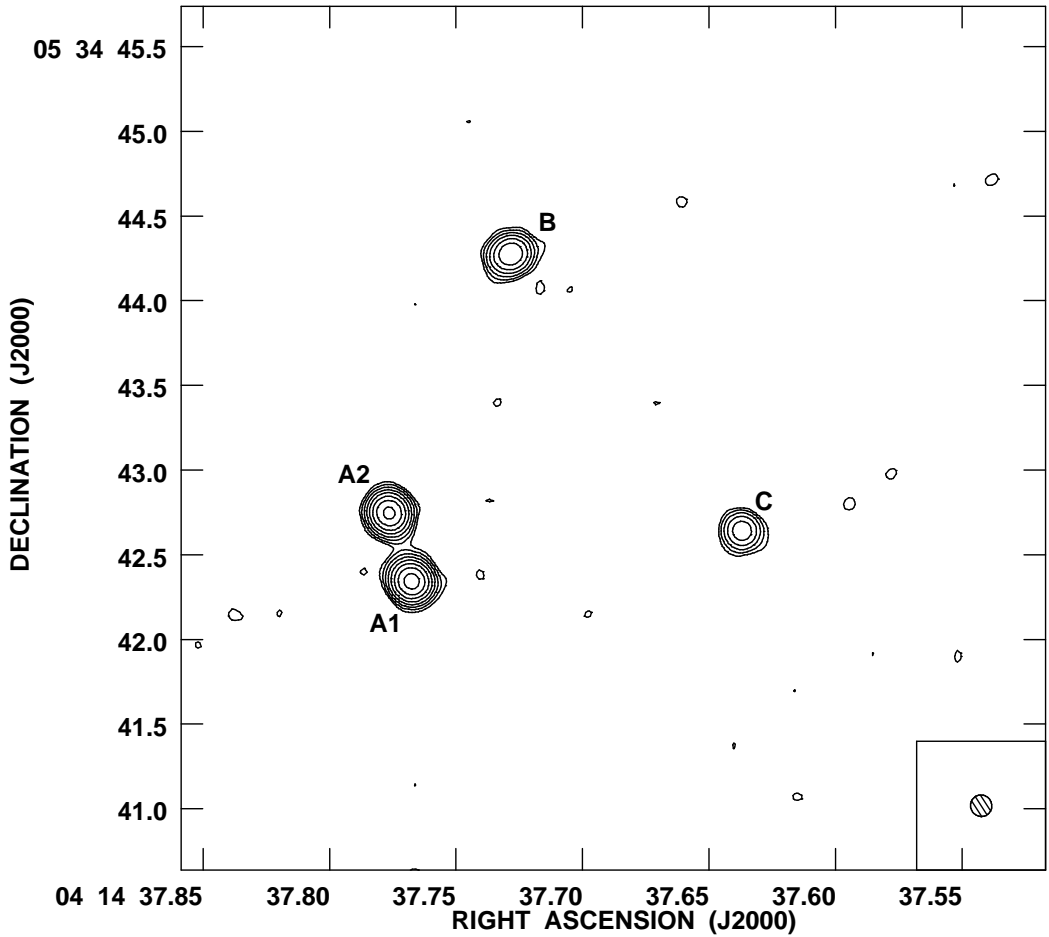}
  \end{center}
\caption{A 15\unit{GHz} (A-array) map of MG0414 produced in the course of the
  present work (observation date 15~Dec~1992).  The contours
  correspond to flux densities of 0.8, 1.6, 3.2, 6.4, 12.8, 25.6, 51.2,
  and 102.4\unit{mJy}.  The circle in the lower right corner
  illustrates the size of the synthesized beam for this observation
  (.15\arcsec$\times$.14\arcsec).}\label{fig:0414map}
\end{center}
\end{figure}

For a time delay measurement we have selected the
gravitational lens MG~0414+0534, a source that was culled from the
MIT--Greenbank survey \cite{bennett86} in a gravitational lens search
\cite{hewitt88} and followed up with optical and radio observations.
The object consists of four bright, highly reddened images with very
similar radio spectra and optical colors.  These characteristics,
along with the morphology and relative brightnesses of the images,
were taken as early evidence that the object is indeed a gravitational
lens \cite{hewitt92a}.  Further study has supported that finding.
Katz and Hewitt \cite*{Katz93a} confirmed that the close double
(components~A1 and A2) is indeed a double as predicted by lens models.
Optical (I-band) observations at the Michigan-Dartmouth-MIT
observatory at Kitt Peak have detected the lensing galaxy at nearly
the expected position \cite{schechter93a}.  Spectroscopy by
Angonin-Willaime \etal\ \cite*{angonin94a} shows that the spectra of
A1+A2 and B are similar.  Infrared spectroscopy by Lawrence \etal\
\cite*{lawrence95b} suggests that the source is a fairly typical
quasar that is highly reddened by dust in the lensing galaxy (but see
Vanderriest \etal\cite*{vanderriest95a} and Annis
\etal\cite*{annis93a} for other views).  A 15\unit{GHz}\ radio image
of MG0414 is displayed in Figure \ref{fig:0414map}.

MG0414 is well suited to relating time delay measurements to angular
diameter distances since the morphology of the system argues for a
relatively simple lensing potential (particularly compared to
QSO B0957+561).  Furthermore, HST, VLA, and VLBI studies of the object are
underway which have revealed structure that may be used to place
constraints on models of the lensing potential
\cite{falco97a,trotter97a,katz96a,ellithorpe95c,patnaik95a}.

\subsection{Lens Models}
\begin{figure}[htb]
\begin{center}
  \begin{minipage}[t]{.49\textwidth}
    \begin{center}\leavevmode\epsfxsize=.95\textwidth
    \epsfbox{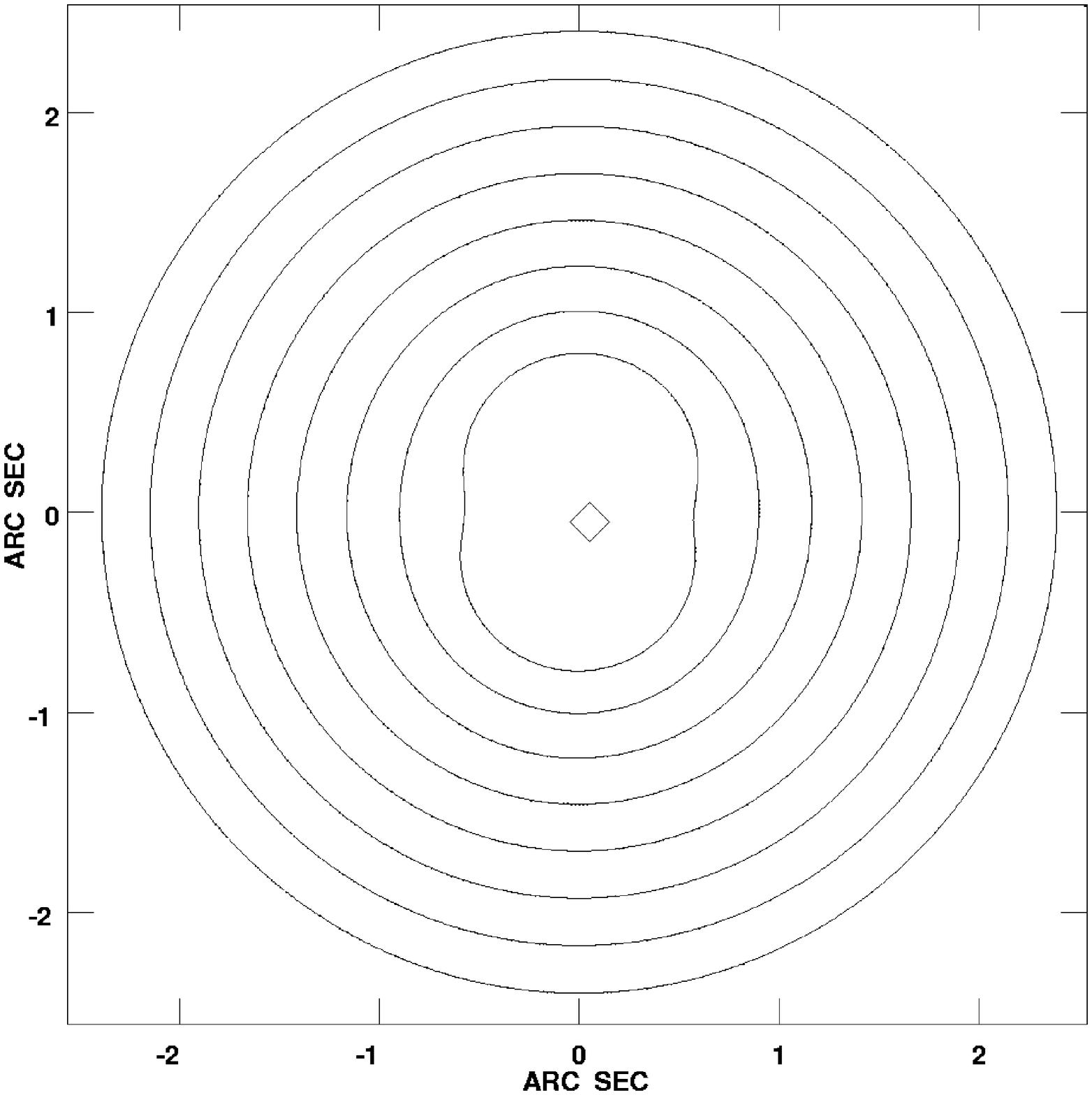}
    \end{center}
  \end{minipage}
  \begin{minipage}[t]{.49\textwidth}
    \begin{center}\leavevmode
    \epsfxsize=.95\textwidth
    \epsfbox{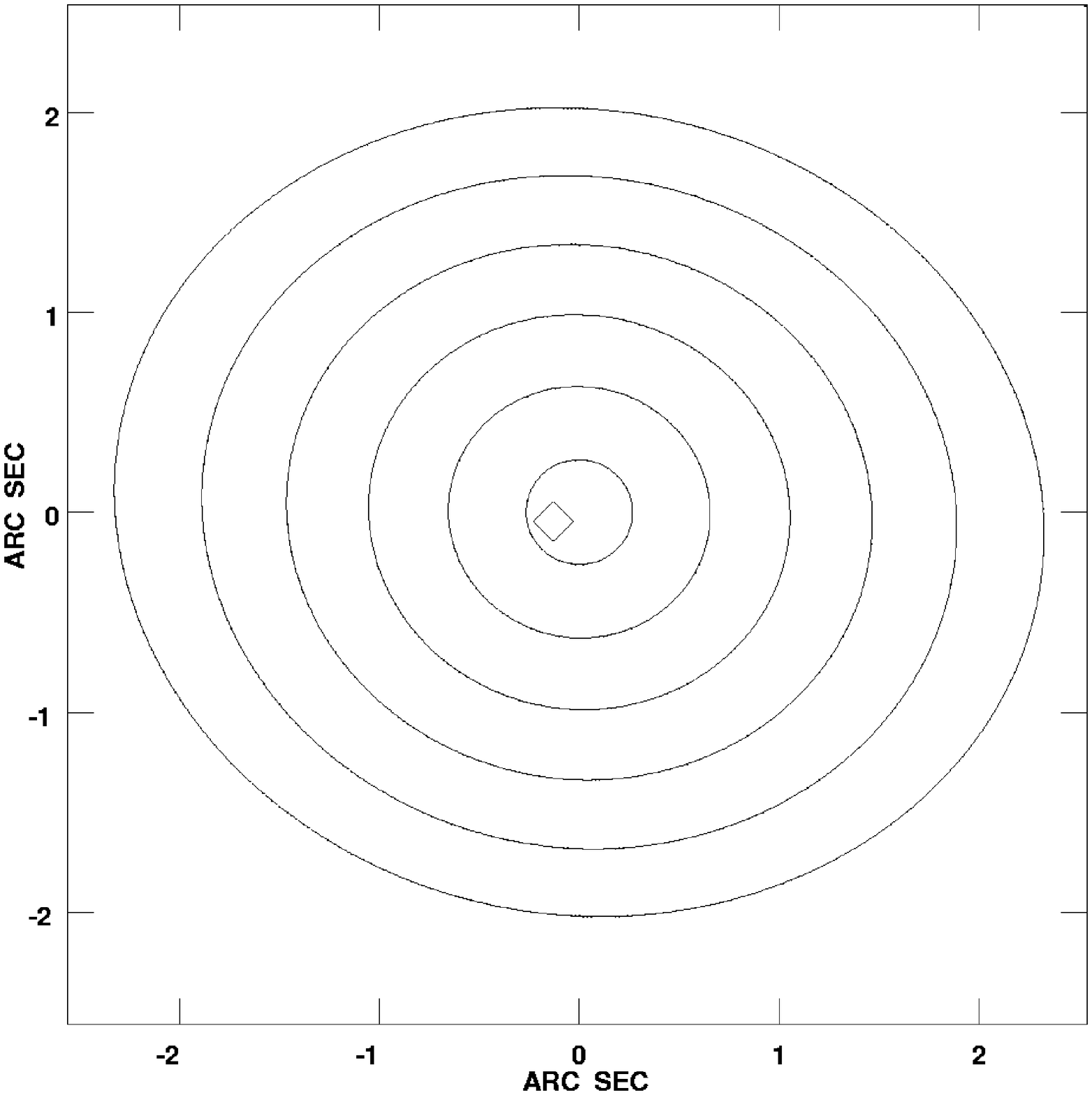}
  \end{center}
  \end{minipage}
\end{center}
    \caption{The gravitational potential for
      Kochanek's model~1 (left) and model~3 (right).  The
      contours are equipotentials which are equivalent (in the weak
      field limit) to lines of constant index of refraction.  The
      diamond signifies our estimated source position.  The central
      detail in model~1 has been suppressed in order to make the
      source position visible.}
    \label{fig:potential}
\end{figure}

Kochanek \cite*{kochanek91a} has fit five simple models of the lensing
potential in MG0414 using {\em only\/} the positions of the images as
constraints for reasons of computational speed and lack of reliable
data on the flux ratios.  The models fall into two classes: those in
which image~B is at a minimum of the time delay surface (models~2, 3,
and 4) and those in which it is at a saddle point (models~1 and 5).
In order to illustrate the difference between the two classes of
models we show Kochanek's model 1 (Singular Isothermal Sphere +
Internal Quadrupole) and model 3 (Singular Isothermal Sphere +
External Quadrupole).  Since Kochanek does not publish the fitted
source positions, we estimate the source position by computing the
source position implied by each observed image.  Images~A1, A2, and B
give consistent estimates of the source position and we take their
average as our estimate ($0.053\arcsec$,$-0.050\arcsec$ from the lens
center for model 1 and $-0.13\arcsec$,$0.046\arcsec$ for model 3).
Figure~\ref{fig:potential} shows the lensing potentials for both
models (from Kochanek's fits) and our estimated
source position.  
Even though the position angles of the potentials differ by
approximately 90\deg, the radial deflections and
therefore the Einstein rings generated by these models are nearly
aligned.

\begin{figure}[htb]
\begin{center}
  \begin{minipage}[t]{.49\textwidth}
    \begin{center}\leavevmode\epsfxsize=.95\textwidth
      \epsfbox{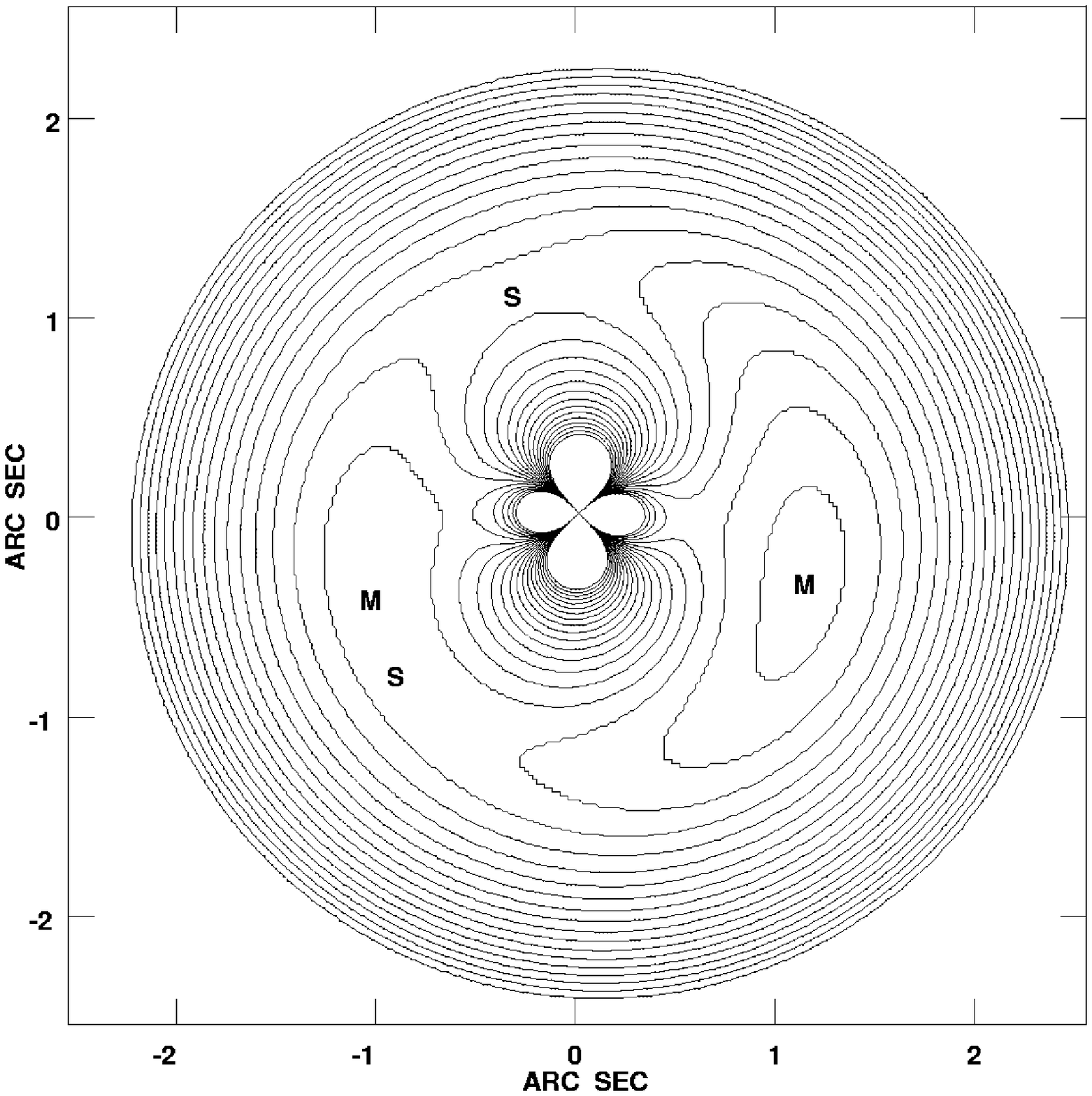}
    \end{center}
  \end{minipage}
  \begin{minipage}[t]{.49\textwidth}
    \begin{center}\leavevmode
    \epsfxsize=.95\textwidth
    \epsfbox{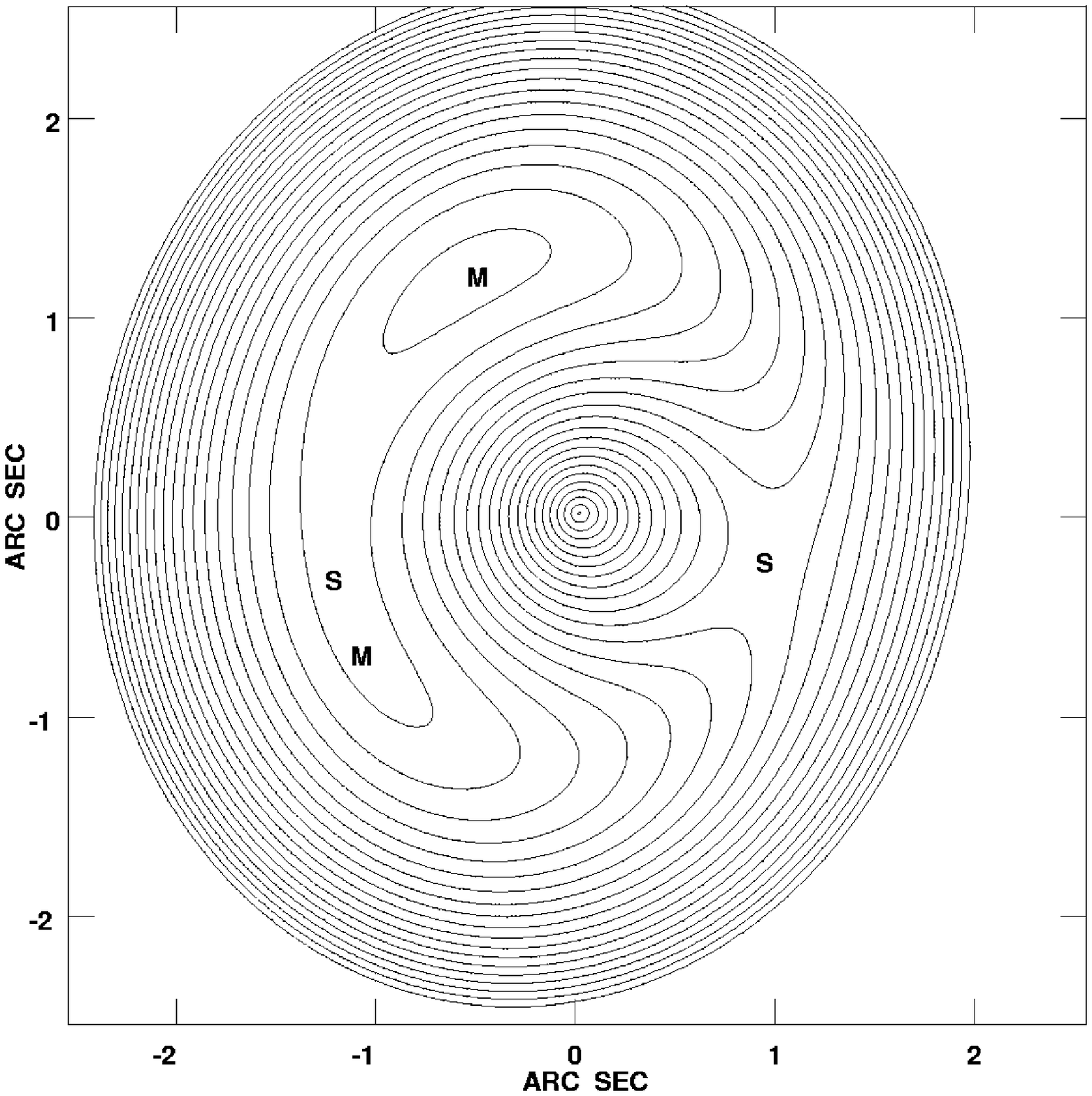}
  \end{center}
  \end{minipage}
\end{center}
  \caption{The time delay surface for Kochanek's model 1(left)
    and model 3(right) using the assumed source positions shown in
    Figure~\protect\ref{fig:potential}.  Extrema (with the exception
    of the central maximum) are labelled ``M'' for minimum and ``S''
    for saddle point.}
    \label{fig:tds}
\end{figure}

Figure~\ref{fig:tds} displays the time delay surfaces corresponding to
the two models.  By the Fermat principle, images are formed at the
extrema (marked) of the time delay surface.  Both models can reproduce
the positions of the images, but they do so very differently: saddle
points and minima are exchanged in the two time delay surfaces.
The B:C flux ratio predicted by Kochanek's models~1 and 5 (.58 and .51
respectively) is inconsistent with our observations where the measured
flux ratio is 2.53.  Models~2, 3, and 4 come closer to the measured B:C flux
ratio (2.2--3.2) but fail to correctly predict the A1:A2 flux ratio.
The very 
large magnification gradient in the neighborhood of A1 and A2
should make the flux ratio of A1:A2 difficult to model.

The models that are consistent with the measured flux ratios predict
that flux variations should be observed first in image~B, following
next in A1 \& A2, and finally in image C.  The model of the system by
Hewitt \etal\ \cite*{hewitt92a} predicts (for $H_0 =
80\unit{km\,s^{-1} Mpc^{-1}}$, $q_0 = 0.5$, empty beam) that images~B
and A2 lead A1 by 12 and 0.3\unit{days} respectively and that image~C
lags A1 by 19\unit{days}.  It should be noted that the model upon
which these estimates are based should be taken to be illustrative
only since it assumes a lens redshift ($z_l = 0.47$) and does not exactly reproduce
the observed geometry and flux ratios of the system.

\section{Observations}
We observed MG0414 at 15\unit{GHz} with the Very Large
Array\VLAfootnote~(VLA) starting on 1992~November~8 in its most
extended (A-array) configuration, continuing through 
the second most extended (B-array) configuration, and ending on
1993~May~3.  The central frequencies for the two observing bands were
the standard VLA U-band values of 14.9649 and 14.9149\unit{GHz},
each with 50\unit{MHz} bandwidth.  Each band was observed in both
senses of circular polarization bringing the total observed bandwidth
to 100\unit{MHz} in each polarization.  There were a total of 63
individual observations each of 0.5 to 1.5 hours duration.  The total
time spanned by the observing program and the sampling interval of the
observations were chosen so that for a reasonable range of the (unknown)
 lens redshift, at least one delay would be detectable.  This
program was scheduled into small interstices in the VLA schedule
\begin{table}[htb]
\begin{tiny}
\begin{center}
\begin{tabular}{rccc||rccc} \hline\hline
&Julian Date & Source Elevation & Synthesized & & Julian Date & Source Elevation & Synthesized \\
\multicolumn{1}{c}{Date} & $- 244000.0$ & (degrees) & Beam
(\arcsec) & \multicolumn{1}{c}{Date} & $- 244000.0$ & (degrees)
& Beam (\arcsec) \\ \hline
1992 NOV 02 & 8928.674 &    13--19 & $.35 \times.23  $ &1993 FEB 01 & No Data    & &  \\                          
NOV 03 & No Data &     &   &                             FEB 03 & 9021.685    & 50--52 & $.48 \times.24  $  \\
NOV 04 & 8930.997 &    28--34 & $.27 \times.25  $  &     FEB 06 & 9024.677    & 50--52 & $.50 \times.30  $  \\
NOV 07 & No Data &    60--62 & $.16 \times.13  $  &      FEB 09 & 9027.690    & 40--44 & $.58 \times.26  $  \\
NOV 13 & 8939.772 &    47--50 & $.15 \times.13  $  &     FEB 12 & 9030.533    & 50--52 & $.45 \times.23  $  \\
NOV 14 & 8940.894 &    53--55 & $.15 \times.13  $  &     FEB 14 & 9032.665    & 40--48 & $.53 \times.38  $  \\
NOV 18 & 8944.738 &    50--52 & $.16 \times.15  $  &     FEB 15 & 9033.673    & 40--44 & $.59 \times.40  $  \\
NOV 23 & 8949.744 &    47--50 & $.22 \times.22  $  &     FEB 18 & 9036.707    & 29--34 & $.56 \times.50  $  \\
NOV 24 & 8950.971 &    22--26 & $.33 \times.20  $  &     FEB 19 & 9037.522    & 50--54 & $.72 \times.33  $  \\
NOV 25 & 8951.864 &    53--55 & $.16 \times.14  $  &     FEB 23 & 9041.704    & 23--33 & $.39 \times.39  $  \\
NOV 26 & 8952.692 &    37--41 & $.15 \times.15  $  &     FEB 27 & 9045.661    & 35--39 & $.43 \times.37  $  \\
NOV 27 & 8953.963 &    25--28 & $.32 \times.20  $  &     MAR 05 & 9052.473    & 50--52 & $.40 \times.39  $  \\
NOV 29 & 8955.707 &    50--52 & $.15 \times.13  $  &     MAR 09 & 9056.357    & 29--34 & $.69 \times.67  $  \\
NOV 30 & 8956.850 &    53--55 & $.16 \times.14  $  &     MAR 15 & 9062.474    & 52--55 & $.39 \times.32  $  \\
DEC 03 & 8959.754 &    52--54 & $.17 \times.17  $  &     MAR 18 & 9065.501    & 54--55 & $.41 \times.33  $  \\
DEC 07 & No Data &     &   &                             MAR 24 & 9070.575    & 40--44 & $.38 \times.35  $  \\
DEC 13 & 8969.814 &    48--50 & $.16 \times.14  $  &     MAR 31 & 9077.535    & 44--48 & $.43 \times.35  $  \\
DEC 15 & 8971.845 &    43--48 & $.15 \times.14  $  &     APR 01 & 9078.539    & 40--48 & $.45 \times.35  $  \\
DEC 16 & 8972.617 &    32--36 & $.14 \times.14  $  &     APR 03 & 9081.334    & 34--41 & $.45 \times.35  $  \\
DEC 24 & 8980.533 &    20--24 & $.35 \times.22  $  &     APR 05 & 9083.476    & 52--54 & $.38 \times.33  $  \\
DEC 27 & 8983.673 &    51--53 & $.17 \times.14  $  &     APR 06 & 9084.494    & 49--52 & $.37 \times.33  $  \\
DEC 29 & 8985.602 &    37--41 & $.14 \times.14  $  &     APR 11 & 9089.516    & 40--44 & $.44 \times.35  $  \\
1993 JAN 04 & 8991.671    & 53--54 & $ .16\times.14  $  &APR 12 & 9090.499    & 45--48 & $.37 \times.34  $  \\
JAN 06 & 8993.634 &    49--53 & $.17 \times.14  $  &     APR 15 & 9093.491    & 45--48 & $.39 \times.34  $  \\
JAN 07 & 8994.699 &    54--55 & $.17 \times.15  $  &     APR 19 & 9097.417    & 54--55 & $.38 \times.34  $  \\
JAN 11 & 8999.514 &    23--30 & $.27 \times.22  $  &     APR 20 & 9098.498    & 44--48 & $.35 \times.33  $  \\
JAN 16 & No Data &     &   &                             APR 22 & 9100.493    & 40--44 & $.38 \times.34  $  \\
JAN 19 & No Data &     &    &                            APR 24 & 9102.343    & 58--59 & $.43 \times.39  $  \\
JAN 21 & 9008.704 &    49--51 & $.16 \times.14  $  &     APR 27 & 9105.315    & 47--49 & $.38 \times.34  $  \\
JAN 24 & 9011.607 &    60--61 & $.17 \times.16  $  &     APR 29 & 9107.286    & 48--52 & $.41 \times.35  $  \\
JAN 30 & 9017.614 &    54--55 & $.42 \times.25  $  &     MAY 01 & 9109.247    & 29--40 & $.36 \times.34  $  \\
&&&& MAY 03 & 9111.378    & 54--55 & $.41 \times.32  $  \\ \hline\hline

\end{tabular}
\end{center}
\end{tiny}

\caption{Journal of Observations}\label{tab:journal}
\end{table}
resulting in observations occurring over a wide range of hour angle
and with no particular sampling pattern (see Table \ref{tab:journal}).
We have found that, in practice, an irregular sampling pattern is
desirable because it minimizes windowing effects in the time delay
analysis.

Some of the observations took place when the source was at very low
elevation.  These data are problematic since the projection of the
baselines of the array and the resulting large synthesized beam makes
deconvolution difficult, and because the atmospheric contribution to
phase errors and flux calibration errors is much larger.

\section{Data Reduction}
The data were reduced using the National Radio Astronomy Observatory's
Astronomical Image Processing System~(AIPS).  After the initial
excision of raw data that were clearly corrupted by interference or
hardware problems, all data sets were processed by AIPS with a ``run
file'' that applied the same processing to each observation.  Complex
antenna gains were determined by observations of J0423$-$013.  Refined
estimates of the antenna phases were obtained by self-calibration
\cite{cornwell89}.  Observations of 3C~48 were used to set the flux
density scale but in order to use as many of the available baselines as
possible, we used a model of the source based on four minutes of
A-array observation kindly provided by C.~Katz.

We used the CLEAN \cite{hogbom74} algorithm implemented within the
AIPS task MX \cite{clark80,schwab84} to deconvolve each image.  CLEAN
is an iterative procedure that, for each iteration, selects the
brightest pixel in the map and subtracts a fraction of it
multiplied by the dirty beam.  As a result, when two point sources are
very close together, components tend to fall spuriously between them.
This effect becomes particularly troublesome for the B-array data in
which the A1--A2 doublet is barely resolved.  Our initial attempts to
clean these data resulted in a single bright source located between
the two images with flux extending from the center towards the true
image positions.  When we instead used the knowledge of the relative
positions of the components as measured from our A-array maps to
constrain CLEAN so that it only attempted to place clean components in
0.12\arcsec~square boxes centered on each component, we found that the
quality of the deconvolution improved as evidenced by comparison of
the B-array maps to the A-array maps.

When two point sources are barely resolved it is difficult to measure
their flux densities independently.  We attempted to use the AIPS task
JMFIT to fit two gaussians to the A1--A2 double and found a strong
covariance between the measured flux densities.  Since we have already
fitted for the flux densities of each component in the image
deconvolution step, we instead measure the flux density of each image
by summing the flux density of the CLEAN components in the box
centered at the image position.  This CLEAN flux density does not
reliably measure the total flux density when the clean components do
not reliably represent the total flux distribution (e.g.~for extended
sources).  Since all four images in MG0414 are nearly point-like,
the CLEAN components contain essentially all of the flux and can be
used directly to get the flux densities.  Our flux density measurement
was compared to the results of JMFIT by creating artificial data sets
with the same {\em uv\/}-sampling as typical data and four gaussian components
with parameters measured from our A-array images.  Gaussian white
noise was added to all visibilities so that the noise in the
image was comparable to the noise measured in the real
images.  The artificial data sets were deconvolved in exactly the same
way as the real data and the resulting images then measured with JMFIT
and by our technique.  The results are comparable for
the case when all images are well resolved (A-array) but in the
B-array data, the A1--A2 pair is marginally resolved and our flux densities
for those images had a variance smaller than those of JMFIT by a factor
of $\sim$50.

Weather conditions can make flux calibration at 15\unit{GHz} very
difficult.  The most important effects are varying absorption due to
the proximity to the water absorption line at 22\unit{GHz} and
wind-induced pointing errors of the individual array antennae.  For
example, we have measured variations in calibrator flux density of a
factor of $\sim 1.5$ over a period of 25\unit{minutes} at the
beginning of a snowstorm.  (The data from that particular observation
are omitted from our analysis.)  Since flux calibration errors can
substantially affect the results of the analyses to be discussed
below, these analyses are carried out using only those observations
made when the wind speed was less than~10\unit{m/s}, the object was
above 30\deg\ elevation, there was no precipitation, and the phase
stability of the atmosphere allowed reliable calibration of the
complex antenna gains.  Of the original 63 observations, three were
lost to equipment failures at the VLA site, calibrations for two were
lost to human error at operations, and one was lost to snow filling
the dishes.  Of the remaining 57 observations, we exclude from
analysis five for high wind, eight for precipitation, two for
atmospheric phase instability, and seven for low elevation, leaving 35
observations.  Table~\ref{tab:fluxes} lists the measured flux
densities for all components for each of the 35 observations.
\begin{table}[htb]
\begin{small}
\begin{center}
\begin{tabular}{ccccc||ccccc} \hline\hline
Julian Date & \multicolumn{4}{c||}{Flux Densities (mJy)} & Julian Date & \multicolumn{4}{c}{Flux Densities (mJy)} \\
 $- 2440000.0$ & A1 & A2 & B& C & $- 2440000.0$ & A1 & A2 & B& C \\ \hline
8930.9968 & 153.8 & 134.6 & 59.8 & 22.6 & 9021.6846 & 154.0 & 137.0 & 59.0 & 22.9 \\
8939.7717 & 163.7 & 146.0 & 66.1 & 22.9 & 9024.6766 & 151.4 & 134.6 & 58.4 & 22.1 \\
8940.8936 & 153.2 & 135.8 & 59.2 & 22.6 & 9030.5334 & 156.7 & 139.1 & 59.0 & 22.4 \\
8944.7380 & 161.7 & 142.2 & 63.7 & 23.1 & 9032.6650 & 151.8 & 136.5 & 57.7 & 22.6 \\
8951.8637 & 148.4 & 131.3 & 58.3 & 21.7 & 9036.7067 & 155.1 & 137.1 & 58.8 & 22.6 \\
8952.6917 & 158.0 & 139.7 & 60.5 & 23.4 & 9045.6612 & 157.7 & 138.8 & 59.6 & 23.1 \\
8955.7073 & 160.8 & 140.8 & 62.7 & 23.3 & 9052.4733 & 154.1 & 133.8 & 58.9 & 20.9 \\
8956.8501 & 152.7 & 136.6 & 60.0 & 22.9 & 9070.5746 & 159.6 & 140.0 & 59.6 & 23.4 \\
8959.7535 & 147.6 & 130.5 & 58.2 & 21.4 & 9077.5345 & 162.4 & 145.0 & 61.0 & 23.7 \\
8971.8451 & 150.8 & 132.7 & 58.3 & 22.5 & 9078.5392 & 160.3 & 143.0 & 61.3 & 22.9 \\
8983.6726 & 150.8 & 133.9 & 59.4 & 22.0 & 9081.3337 & 149.4 & 129.9 & 55.1 & 21.4 \\
8991.6712 & 149.6 & 133.0 & 57.3 & 22.3 & 9083.4761 & 144.5 & 127.0 & 54.2 & 20.7 \\
8993.6338 & 163.8 & 145.5 & 62.8 & 24.4 & 9093.4906 & 152.1 & 132.8 & 56.9 & 21.5 \\
8994.6986 & 148.8 & 131.7 & 57.7 & 22.2 & 9098.4980 & 152.5 & 133.3 & 56.8 & 21.9 \\
9008.7038 & 148.1 & 132.9 & 57.0 & 21.1 & 9100.4928 & 158.8 & 139.4 & 59.2 & 23.5 \\
9011.6065 & 146.9 & 130.4 & 58.8 & 22.0 & 9105.3149 & 155.1 & 139.2 & 60.5 & 22.7 \\
9017.6137 & 151.5 & 133.8 & 57.6 & 22.0 & 9107.2857 & 156.1 & 137.2 & 59.7 & 22.8 \\
&&&&                                    & 9111.3782 & 149.2 & 131.7 &56.8 & 22.1 \\ \hline\hline
\end{tabular}
\end{center}
\end{small}

\caption{Table of measured flux densities (in mJy).}\label{tab:fluxes}
\end{table} 

\begin{figure}[htb]
\epsfxsize=\textwidth
  \begin{center}\leavevmode
   \epsfbox{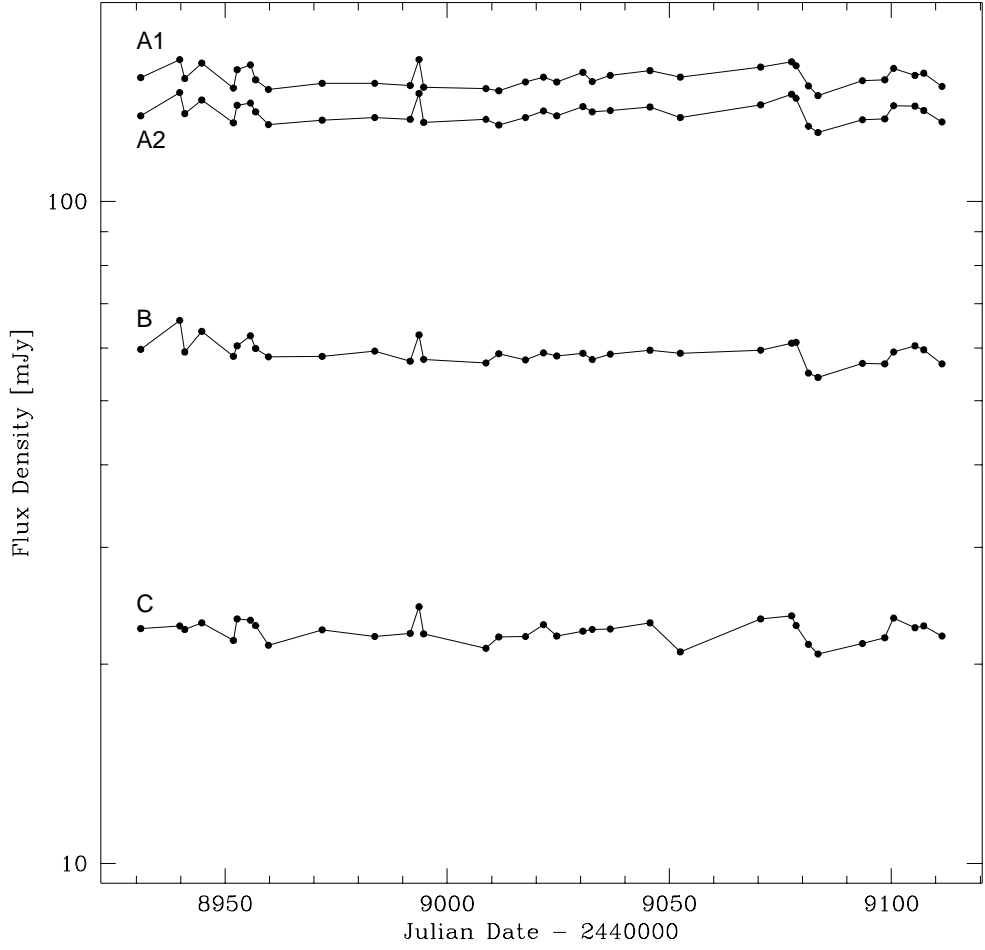}
  \end{center}
  \caption{Light curves for all components shown on a log scale so
    that shifting on the vertical axis is equivalent to scaling.  }
  \label{fig:loglc}
\end{figure}

Figure~\ref{fig:loglc} displays the measured light curves for all four
images in MG0414.  The variability is still clearly dominated by the
flux calibration errors caused by variable atmospheric water content,
differences in source and calibrator elevations, flux calibrator
modelling errors, and weather conditions.  Even with these effects,
the RMS variability in the resulting four light curves (without any
adjustment for possible source variability) is only 3.3--3.9\%.  Since
this includes both source variability and flux calibration errors, the
relative flux calibration is at least this good over the whole
data set.

\section{Analysis}
\subsection{Variability in the Source}\label{sec:var}
\begin{figure}[htb]
\epsfxsize=\textwidth
  \begin{center}\leavevmode
   \epsfbox{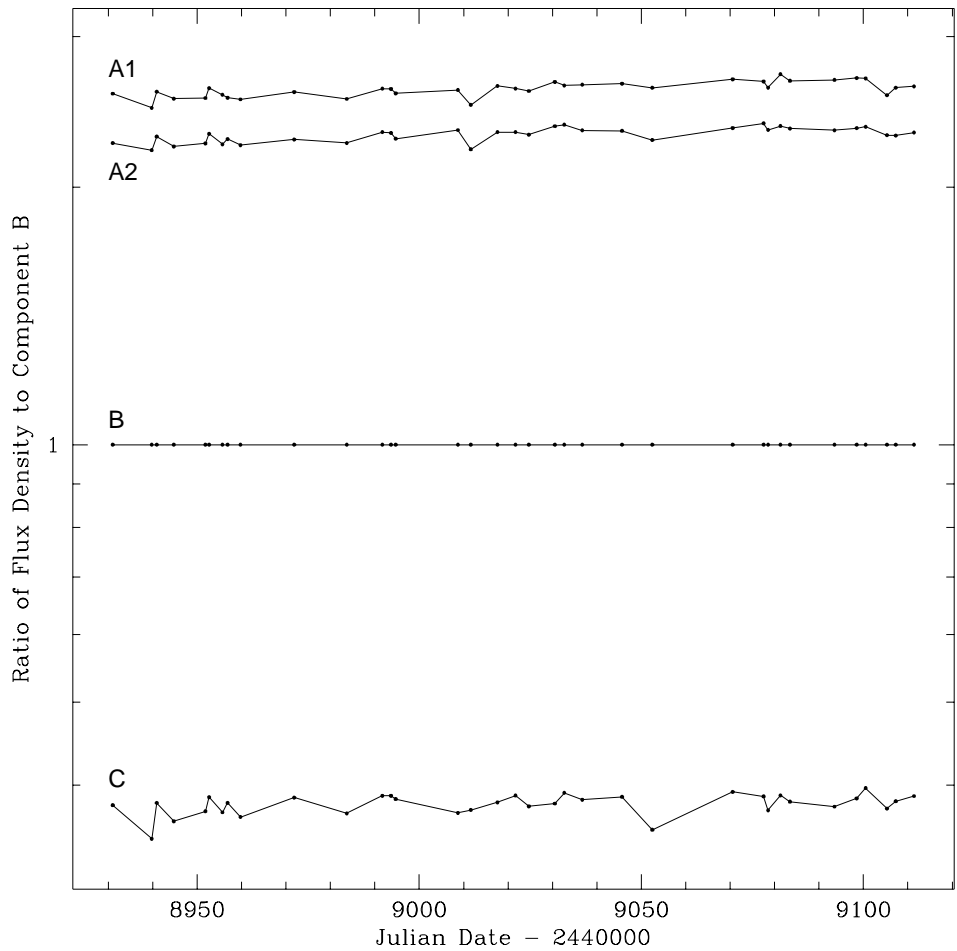}
  \caption{Time history of the ratio of the flux density of each image
    to image B.  Shown on a log scale so that shifting on the vertical
    axis is equivalent to scaling.}
  \label{fig:ratiolc}
  \end{center}
\end{figure}

Source variability is prerequisite to the determination of a time
delay but the uncertainties in the flux calibration obscure any evidence
for source variability in the light curves of Figure~\ref{fig:loglc}.
In order to determine if there is source variability, we compare the
variations in the ratios of various images.  Figure~\ref{fig:ratiolc}
displays ``ratio light curves'' in which the flux density of each
component is normalized to component B.  
For comparison, we generated an artificial
data set with the same {\em uv\/}-sampling as the real data, the average fluxes
for the four components of MG0414 that we measure, and no source
variability (created by by the AIPS task UVMOD).  An estimate of the
RMS noise (per visibility) was obtained from the phase calibrator
observations in the real data and gaussian noise of this amplitude was
added to the visibilities in the artificial data.  This complete
artificial data set was then processed with the same run-file as the
real data.  Table~\ref{tab:fluxratios} displays the RMS variability in
each flux ratio for both the real data and the artificial data
described above.  These results demonstrate that uncertainties in the
map computation and flux estimation procedures are comparable to the
uncertainties in the calibration of the overall flux density scale.
Thus, there is no evidence here that any source variability has been detected.

\begin{table}[htb]
\begin{center}
\begin{tabular}{ccc} \hline
Ratio & Real Data   & Artificial Data \\ \hline\hline
A1:A2 & 0.9\% & 1.4\% \\
A1:B  & 2.1\% & 1.9\% \\
A2:B  & 1.9\% & 1.4\% \\
A1:C  & 2.2\% & 2.6\% \\
A2:C  & 2.1\% & 2.8\% \\
B:C   & 2.9\% & 2.7\% \\\hline
\end{tabular}
\end{center}
\caption{Measured RMS variability in image flux density ratios for the
  data reported here and an artificial data set with similar instrumental
  properties.}\label{tab:fluxratios}
\end{table} 

Another measure of variability is the first order structure
function \cite{simonetti85},
\[ V(\tau) = \frac{1}{2}\left<\left[s(t) - s(t+\tau)\right]^2\right>\]
where $s(t)$ is the image flux density at time $t$, $\tau$ is the
difference in time between two observations, and $\left<\,\right>$
indicates the average.)  
We fit the structure function of the log of the
radio data in decibels (referenced to 1\unit{Jy}), adopting a
structure function of the form
\[ V(\tau) =C \tau^{1.0}.\]
We expect that the structure function of instrumental errors will be
that of white noise, so we take the power law part of the structure
function as an estimate of the maximum possible variability in the
source and find
\[ V(\tau) = (0.000139\unit{\frac{decibels}{day}}) \tau.\]
This implies that the signal-to-noise ratio for a single measurement
of a change in flux density on time scales of interest
($\sim$20\unit{days}) is at most $\sim$0.3.

\subsection{Time Delays and Correlated Variability}
The data presented above provide no evidence that we have detected
variability in MG0414.  However, it is possible that some of the
variability in the data is due to the quasar and that such ``real''
variability would manifest itself as itself in the form of a
correlation between flux density measurements separated by the lens
time delay.  To investigate this, we computed discrete correlation
functions \cite{edelson88a} for pairs of our light curves.  No
evidence for correlations was found in this analysis.

\begin{figure}[htb] \epsfxsize=\textwidth
  \begin{center}\leavevmode
   \epsfbox{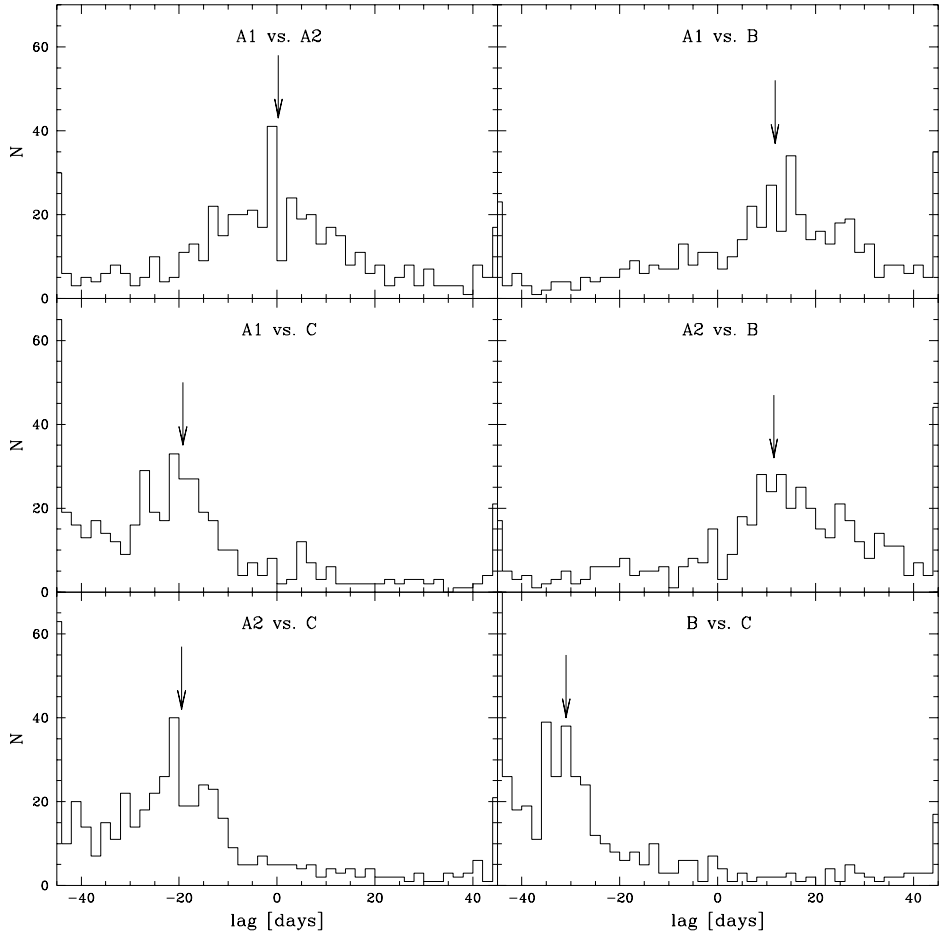}
  \end{center}
  \caption{Histograms of the absolute minimum of the \prhchi\ on the
    interval $[-45,45]$ for 500 Monte Carlo trials assuming the presence
    of a gravitational lens.  The model delay is
    marked with an arrow in each case.  The bin on the left edge of
    the B vs. C plot is off-scale, its value is 121.  }
  \label{fig:montehist}
\end{figure}

A time delay detection technique with greater statistical efficiency
than cross-correlation techniques has been developed by Press,
Rybicki, and Hewitt \cite*{prhchi2,prhchi1}; see also \cite{rybicki92}.  Here we
examine their $\chi^2$ statistic (hereafter \prhchi).  To apply their
method, one assumes (or better, fits for) a flux ratio between two
images and then, for each trial delay, creates a combined light curve
that consists of the union of the two light curves with one of them
shifted and scaled relative to the other.  For each trial delay, we
ask how likely (in a $\chi^2$ sense) the combined light curve is to
have originated from a process whose variability is described by the
assumed structure function.   We have performed Monte Carlo
simulations by generating 500 trials using the sampling of the data
from this work, the delays predicted by the Hewitt \etal\
\cite*{hewitt92a} models, the noise measured from the data (assumed to
be white gaussian), and assuming the source variability to be
generated by a gaussian process with the same structure function as
that estimated from the data.  For comparison, we also performed the
same simulation assuming no gravitational lens (i.e. every image is an
independent source of gaussian noise again with the same structure
function as that estimated from the data but uncorrelated with the
other images).  Figure~\ref{fig:montehist} displays a histogram of the
absolute minimum of the \prhchi\ curve on the interval [-45,45] for
every image pair in the presence of the modelled gravitational lens.
The simulations show that given the quality and quantity of the
available data and the small (if present) variability of the source,
extraction of a time delay is not possible by this technique.  Only
22--31\% of the measurements are within $\pm5\unit{d}$ of the
simulated delay.  However, it is clear that the technique is detecting the
presence of correlated variability since the correlated simulations
show 3--6 times more measurements within 5\unit{d} of the simulated
delay than the uncorrelated (no gravitational lens) simulations.


Even given the low signal-to-noise ratio of the present data, the
\prhchi\ method is interesting since it uses all available lag pairs
and is thus a very sensitive way to measure correlated variability.
Given sufficiently large data sets, one can expect to detect correlated
variability even when the signal-to-noise ratio for a single lag pair
is less than unity \cite{hewitt95a}.

\begin{figure}[htb]
\epsfxsize=\textwidth
  \begin{center}\leavevmode
   \epsfbox{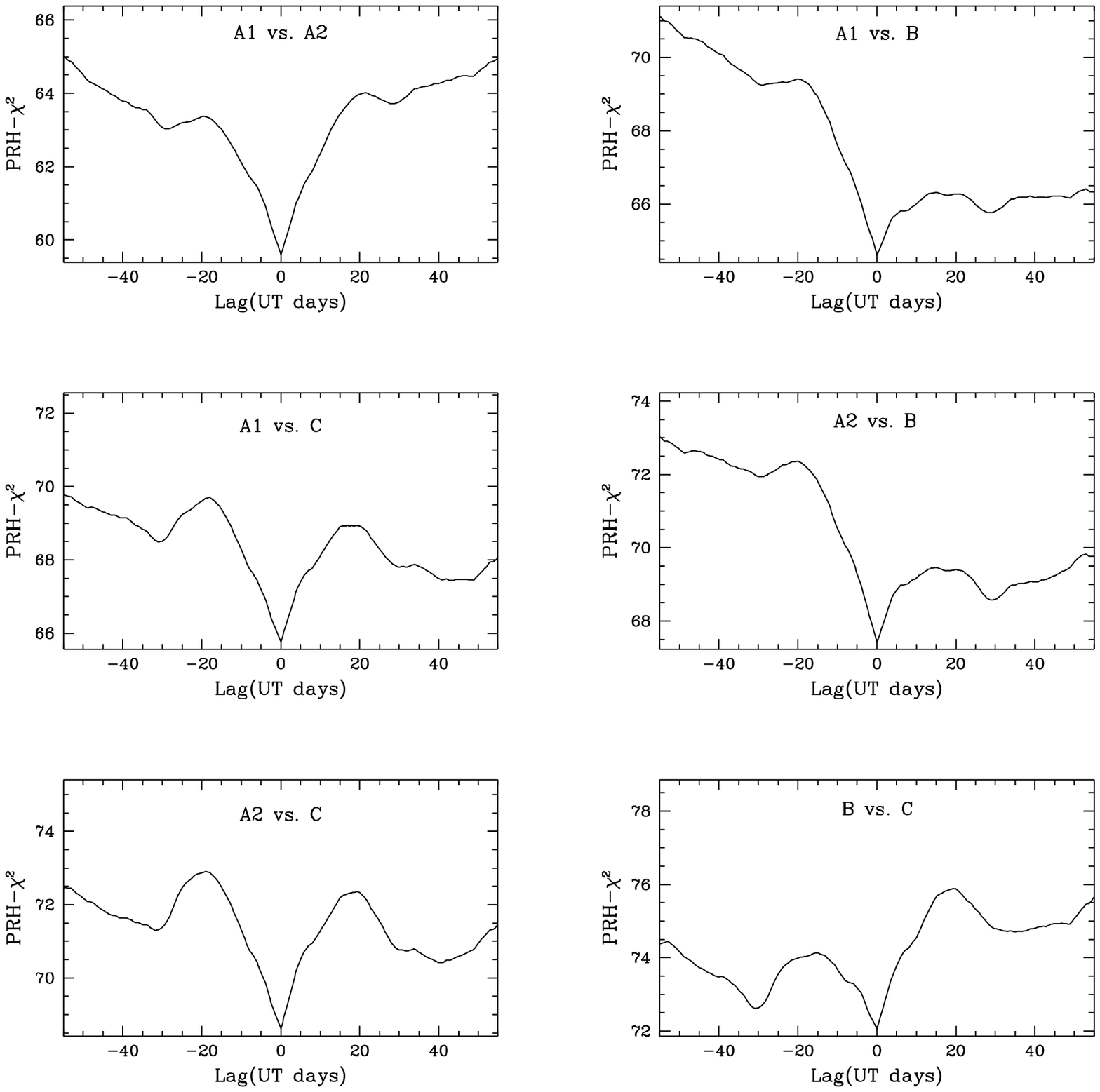}
  \end{center}
  \caption{PRH-$\chi^2$ curves for all pairs of components (69 degrees
    of freedom).  Note the prominent feature at zero lag due to flux
    calibration errors.  Also note that some of the curves,
    particularly those in the right column exhibit a strong preference
    for a particular {\em sign\/} of the delay.  For ease of
    comparison, the curves are all plotted with the same vertical
    scale.  }
  \label{fig:prhchiall}
\end{figure}

From the measured MG0414 data, one can compute a distinct \prhchi\ curve for
every pair of images using the average flux ratio as an estimate of
the true flux ratio.  The six curves computed from the present data are
displayed in Figure~\ref{fig:prhchiall}. %
It is clear from examining the \prhchi\ curves that there is a strong
correlation between the light curves at zero lag due to unmodelled
errors in the absolute flux calibration and that none of them presents
a convincing argument for any particular time delay.  This is not
surprising.  The signal-to-noise ratio of 0.3 for the variability as
measured in a single lag pair (see Section \ref{sec:var} above) can be
compared to the case of Hewitt {\em et al.'s\/}\cite*{hewitt95a}
MG1131+0456 data for which the same ratio is 0.7.  Their simulations
of time delay observations in MG1131 show that 50 to 100 flux
measurements are required for a measurement of the time delay.
Clearly this study of MG0414 suffers because there are too few
measured points, given the small signal-to-noise ratio.  It is
interesting to note that in the cases where the \prhchi\ curves show a
particular preference for the {\em sign\/} of the delay, it is
consistent with the predictions of Kochanek's models 2, 3, and 4
\cite{kochanek91a} and the Hewitt \etal\ \cite*{hewitt92a} model.

\section{Conclusions}

Measurement of a time delay from these data is difficult at best
because the variability is very small, the small size of the data set
gives few lag pairs, and the zero lag correlation caused by errors in
flux calibration dominates the \prhchi\ curves.  One could, in
principle, modify the statistical model of the data used in the
$\chi^2$ analysis to account for the zero-lag correlation of the flux
density calibration errors.  However, given the limited expectation of
a determination of a time delay we postpone such analysis until a
better data set is available.  Our current analysis does not provide a
convincing case for source variability or for any particular value of
a time delay in MG0414.

Improved experiments for measuring this time delay might use one or
more of the following strategies: obtaining a larger data set,
observing at a lower frequency where flux calibration is more precise,
or (conversely) observing at a higher frequency where the source is
more likely to be strongly variable.

\section{Acknowledgements}
We gratefully acknowledge useful advice and comments from 
E.W.~Bertschinger, R.A.~Perley,
S.R.~Conner, R.C.~Walker, W.H.~Press, and D.~Wunker.  The observations
described herein were greatly assisted by the efforts of the VLA
staff.  This work was supported by a David and Lucile Packard
Fellowship in Science and Engineering, a National Science Foundation
Presidential Young Investigator Award, and the M.I.T. Class of~1948.

\FloatBarrier

\newpage 
\bibliographystyle{astro}
\bibliography{radio}
\end{document}